\documentclass[
superscriptaddress,aip,apl,reprint,twocolumn,amsmath,amssymb,showpacs]{revtex4-2}

\usepackage{graphicx}  
\usepackage{mathptmx}
\DeclareMathAlphabet{\mathcal}{OMS}{cmsy}{m}{n}
\DeclareSymbolFont{largesymbols}{OMX}{cmex}{m}{n}
\usepackage{epstopdf}
\usepackage{dcolumn}   
\usepackage{bm}        
\usepackage{amssymb}   
\usepackage{amsmath}   
\usepackage{amsthm}
\usepackage{chemarrow}
\usepackage{color}
\usepackage{mathrsfs}
\usepackage{float}
\usepackage{bbold}
\usepackage{bbm}
\usepackage[a4paper,colorlinks=true,
linkcolor=blue,citecolor=blue,
pdfauthor={ },
pdftitle={ },
pdfsubject={ },
pdfkeywords={ }]{hyperref}

\theoremstyle{plain}
\newtheorem*{theorem*}{Theorem}

\begin{document}


\title{SAPPHIRE: Search for exotic parity-violation interactions with quantum spin amplifiers}

\date{\today}

\author{Yuanhong Wang}
\email[]{These authors contributed equally to this work}
\affiliation{
CAS Key Laboratory of Microscale Magnetic Resonance and School of Physical Sciences, University of Science and Technology of China, Hefei, Anhui 230026, China}
\affiliation{
CAS Center for Excellence in Quantum Information and Quantum Physics, University of Science and Technology of China, Hefei, Anhui 230026, China}

\author{Ying Huang}
\email[]{These authors contributed equally to this work}
\affiliation{
CAS Key Laboratory of Microscale Magnetic Resonance and School of Physical Sciences, University of Science and Technology of China, Hefei, Anhui 230026, China}
\affiliation{
CAS Center for Excellence in Quantum Information and Quantum Physics, University of Science and Technology of China, Hefei, Anhui 230026, China}

\author{Chang Guo}
\affiliation{
CAS Key Laboratory of Microscale Magnetic Resonance and School of Physical Sciences, University of Science and Technology of China, Hefei, Anhui 230026, China}
\affiliation{
CAS Center for Excellence in Quantum Information and Quantum Physics, University of Science and Technology of China, Hefei, Anhui 230026, China}

\author{Min Jiang}
\email[]{dxjm@ustc.edu.cn}
\affiliation{
CAS Key Laboratory of Microscale Magnetic Resonance and School of Physical Sciences, University of Science and Technology of China, Hefei, Anhui 230026, China}
\affiliation{
CAS Center for Excellence in Quantum Information and Quantum Physics, University of Science and Technology of China, Hefei, Anhui 230026, China}

\author{Xiang Kang}
\affiliation{
CAS Key Laboratory of Microscale Magnetic Resonance and School of Physical Sciences, University of Science and Technology of China, Hefei, Anhui 230026, China}
\affiliation{
CAS Center for Excellence in Quantum Information and Quantum Physics, University of Science and Technology of China, Hefei, Anhui 230026, China}

\author{\mbox{Haowen Su}}
\affiliation{
CAS Key Laboratory of Microscale Magnetic Resonance and School of Physical Sciences, University of Science and Technology of China, Hefei, Anhui 230026, China}
\affiliation{
CAS Center for Excellence in Quantum Information and Quantum Physics, University of Science and Technology of China, Hefei, Anhui 230026, China}

\author{\mbox{Yushu Qin}}
\affiliation{
CAS Key Laboratory of Microscale Magnetic Resonance and School of Physical Sciences, University of Science and Technology of China, Hefei, Anhui 230026, China}
\affiliation{
CAS Center for Excellence in Quantum Information and Quantum Physics, University of Science and Technology of China, Hefei, Anhui 230026, China}

\author{Wei Ji}
\affiliation{Helmholtz-Institut, GSI Helmholtzzentrum f{\"u}r Schwerionenforschung, Mainz 55128, Germany}
\affiliation{Johannes Gutenberg University, Mainz 55128, Germany}

\author{Dongdong Hu}
\affiliation{
State Key Laboratory of Particle Detection and Electronics, University of Science and Technology of China, Hefei, Anhui 230026, China}

\author{Xinhua Peng}
\email[]{xhpeng@ustc.edu.cn}
\affiliation{
CAS Key Laboratory of Microscale Magnetic Resonance and School of Physical Sciences, University of Science and Technology of China, Hefei, Anhui 230026, China}
\affiliation{
CAS Center for Excellence in Quantum Information and Quantum Physics, University of Science and Technology of China, Hefei, Anhui 230026, China}

\author{Dmitry Budker}
\affiliation{Helmholtz-Institut, GSI Helmholtzzentrum f{\"u}r Schwerionenforschung, Mainz 55128, Germany}
\affiliation{Johannes Gutenberg University, Mainz 55128, Germany}
\affiliation{Department of Physics, University of California, Berkeley, CA 94720-7300, USA}

\maketitle
\noindent
\textbf{Quantum sensing provides sensitive tabletop tools to search for exotic spin-dependent interactions beyond the Standard Model,
which have attracted great attention in theories and experiments\cite{demille2017probing}.
Here we develop a technique based on quantum Spin Amplifier\cite{jiang2021search} for Particle PHysIcs REsearch (SAPPHIRE) to resonantly search for exotic interactions, specifically parity-odd spin-spin interactions.
The present technique effectively amplifies the pseudomagnetic field generated by exotic interactions by a factor of about 200 while being insensitive to spurious external magnetic fields.
Our studies, using such a quantum amplification technique, open the doors to exploring the parity-violation interactions mediated by $Z'$ bosons\cite{dobrescu2006spin,fadeev2019revisiting} in the challenging parameter space (force range between $3\,\text{mm}$ and $0.1\,\text{km}$) and set the most stringent constraints on $Z'$-mediated electron-neutron couplings, significantly improving previous limits\cite{antypas2019isotopic} by up to five orders of magnitude.
Moreover, our bounds on $Z'$-mediated couplings between nucleons reach into a hitherto unexplored parameter space (force range below $1\,\text{m}$), complementing the existing astrophysical and laboratory studies\cite{dobrescu2006spin,vasilakis2009limits,hunter2013using}.
}

Numerous theories have predicted the existence of exotic interactions beyond the Standard Model that are distinct from the four known interactions~\cite{moody1984new,dobrescu2006spin}.
These exotic interactions were first predicted to be mediated by hypothetical spin-0 bosons such as axions and axion-like particles~\cite{peccei1977cp,kim2010axions},
whose existence may explain many central puzzles in physics,
including the strong charge-parity problem~\cite{peccei1977cp,kim2010axions} and the nature of dark matter~\cite{jiang2021search,duffy2009axions}.
Subsequently, the interaction mediators were further extended to other new gauge bosons such as massive spin-1 bosons and paraphotons~\cite{dobrescu2006spin},
resulting in sixteen types of potentials under the frame of quantum field theory.
A series of astrophysical and laboratory searches are presented to constrain such exotic interactions~\cite{safronova2018search,demille2017probing}.
Since Lee and Yang first discussed the use of gravitational equivalence tests (E{\"o}tv{\"o}s-type experiments) to constrain a long-range interaction~\cite{lee1955conservation}, recent progress in exploring a broad range of exotic interactions has been enabled by advances in precision techniques~\cite{su2021search,wang2022limits,ji2018new,chu2020search,arvanitaki2014resonantly,vasilakis2009limits,bulatowicz2013laboratory,rong2018constraints,yan2013new,hunter2013using,hunter2014using,piegsa2012limits}.

After parity violation was discovered in beta-decay of $^{60}$Co~[\onlinecite{wu1957experimental}], various searches for parity violation are performed, including different parity non-conservation tests\,\cite{antypas2019isotopic,safronova2018search,dzuba2022long} and static electric dipole moment searching experiments (see a recent review~\cite{chupp2019electric} and references therein).
The manifestation of the parity-violation nature of exotic interactions kindles an interest in the search for parity-odd interactions, as the study on such exotic interactions has expanded the area of parity-violation research, which could be a harbinger of potential new physics effects beyond the Standard-Model expectations. 
In the present study, we investigate the parity-odd spin-spin (POSS) interaction induced by the exchange of hypothetical spin-1 $Z'$ bosons.
Due to the potential loopholes left by astrophysical observations on the constraints of this interaction~\cite{dobrescu2006spin,jain2006evading}, it would be valuable to perform direct tests with tabletop experiments.
However, the $Z'$-mediated POSS interaction has been explored by only a few laboratory experiments~\cite{hunter2013using,vasilakis2009limits,antypas2019isotopic}.
In particular, it remains challenging to extract the sought-after signal at short (meter scale and shorter) distances due to interference from usual magnetic interactions.

\begin{figure}[t]  
	\makeatletter
	\def\@captype{figure}
	\makeatother
	\includegraphics[scale=1.1]{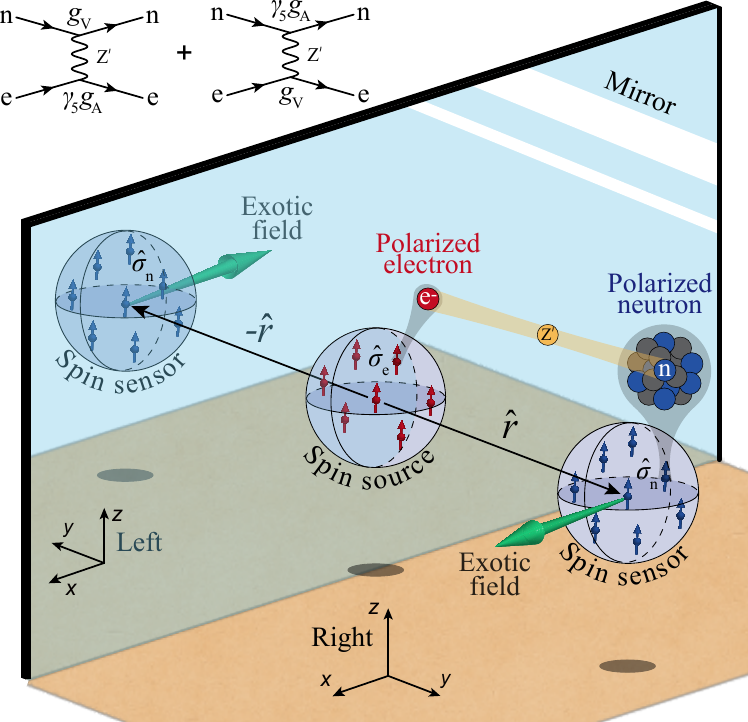}
	\caption{Illustration of parity-odd spin-spin (POSS) interaction.
    The exotic interaction between electrons and neutrons involving the exchange of $Z'$ bosons can be described by the Feynman diagrams~\cite{dobrescu2006spin}.
    As a parity-odd interaction, the induced exotic field  changes sign under a parity transformation as shown in the mirror.
    }
	\label{figure1}
\end{figure}

We focus on the exotic POSS interaction between electrons and neutrons (Fig.\,\ref{figure1}), whose corresponding potential $V_{11}$ following the notation in Refs.\,[\onlinecite{dobrescu2006spin},\onlinecite{fadeev2019revisiting}] is given as ($\hbar=c=1$)
\begin{flalign}
      V_{11}=-f_{11}\left[(\hat{\sigma}_n\times\hat{\sigma}_e)\cdot\hat{r}\right]\left(\frac{1}{\lambda r}+\frac{1}{r^2}\right)\frac{e^{-r/\lambda}}{4\pi m_e},
\label{v11}
\end{flalign}
where $\hat{\sigma}_e$ ($\hat{\sigma}_n$) is electron (neutron) unit spin vector, $\hat{r}$ is the unit vector directed from electrons to neutrons, $r$ is the distance between them, $\lambda$ is the force range and $f_{11}$ is the dimensionless coupling constant.
The potential can be generated by the exchange of an ultralight spin-1 boson $Z'$\,[\onlinecite{dobrescu2006spin},\onlinecite{fadeev2019revisiting}] whose mass $m_{Z'}=\lambda^{-1}$ determines the force range of the interaction.
Experimental searches for exotic spin-spin interactions are typically performed by positioning two polarized objects in close proximity to each other,
with one acting as a magnetic detector (``spin sensor'') to measure the exotic field produced by the other polarized object (``spin source'').
The POSS interaction induces an energy shift of the nuclei in the spin sensor $\Delta E=V_{11}$,
which is due to a pseudomagnetic field $\textbf{B}_{11}\hat{\sigma_n}=-V_{11}/\mu_{\text{N}}$
with $\mu_{\text{N}}$ being the magnetic moment of nuclei.
It is worthy noting that the exotic field $\textbf{B}_{11}\propto -\hat{\sigma}_e\times \hat{r}$ acted on the mirror symmetrical spin sensor is in the opposite direction to the one before parity transformation (Fig.~\ref{figure1}),
illustrating that the parity of the $V_{11}$ interaction is broken.

\begin{figure}[t]  
	\makeatletter
	\def\@captype{figure}
	\makeatother
	\includegraphics[scale=1.15]{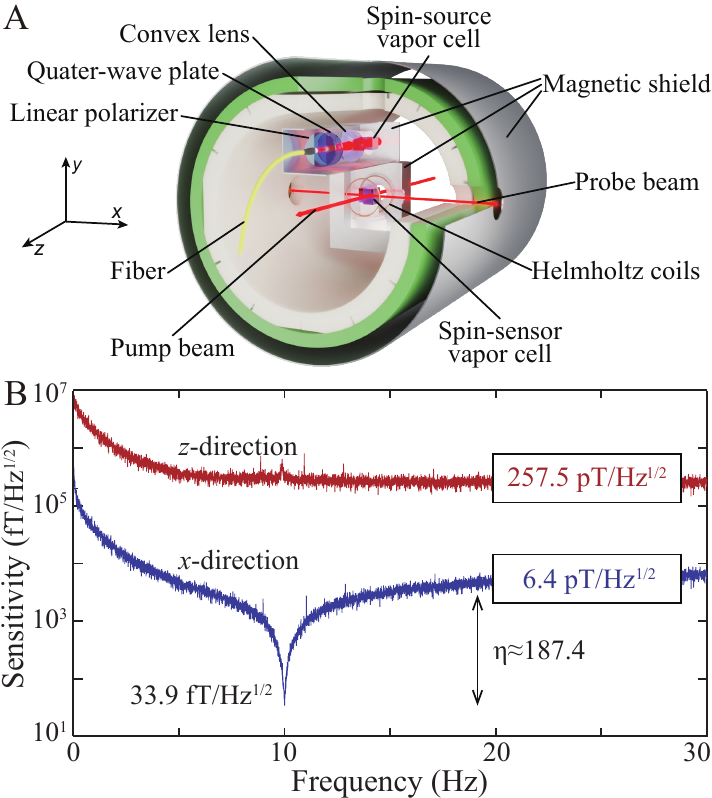}
	\caption{Spin amplification setup and magnetic-field sensitivity.
	(A) The $^{87}$Rb atoms (the spin source) contained in a vapor cell with N$_2$ buffer gas are polarized with a circularly polarized laser beam.
	The spin-sensor cell containing enriched $^{129}$Xe, buffer gas N$_2$ and a droplet of $^{87}$Rb is placed below the spin source.
	A circularly polarized beam along $\hat{z}$ and a linearly polarized beam along $\hat{x}$ are used to, respectively, pump and probe the spin-sensor.
	(B) The sensitivity of the spin amplifier along $\hat{x}$ is measured to be $33.9$\,fT/Hz$^{1/2}$ at the resonance frequency and $6.4$\,pT/Hz$^{1/2}$ off resonance.
	In contrast, the magnetic sensitivity along $\hat{z}$ is only $257.5$\,pT/Hz$^{1/2}$.
	}
	\label{figure2}
\end{figure}

In our experiment, the spin sensor uses polarized neutron spins of isotopically enriched polarized $^{129}$Xe gas and the spin source uses polarized electron spins of $^{87}$Rb atoms.
As shown in Fig.\,\ref{figure2}A, the spin source is positioned $50.7$\,mm above the center of the spin-sensor vapor cell.
The corresponding number of polarized $^{87}$Rb electron spins in the spin source is calibrated about $(2.1\pm0.2)\times10^{14}$ (see Methods and Supplementary Section I for details).

A quantum spin amplifier~\cite{jiang2021search,jiang2021floquetspin} is introduced to resonantly amplify the effect of the exotic field $\textbf{B}_{11}$.
The spin amplifier employs spatially overlapping hyperpolarized long-lived $^{129}$Xe nuclear spins and optically-pumped polarized $^{87}$Rb spins (Supplementary Section II.A).
The $^{129}$Xe spins are polarized by spin-exchange collisions with polarized $^{87}$Rb spins~\cite{jiang2021search,jiang2021floquetspin}. 
A constant field $B_0\hat{z}$ is applied in the sensor, along with an oscillatory transverse field $\textbf{B}_\text{ac}$ to be measured.
When the oscillation frequency of $\textbf{B}_\text{ac}$ matches the $^{129}$Xe Larmor frequency, the polarized $^{129}$Xe spins are tilted away from their polarization axis, and generate a transverse oscillatory magnetization.
Due to Fermi-contact interactions~\cite{jiang2021floquetspin,jiang2021search}, an effective field $\textbf{B}^n_\text{eff}$ generated by the transverse magnetization is read out by the embedded $^{87}$Rb spins that act as an $in$-$situ$ magnetometer, as shown in Expended Data Fig\,\ref{figure3}A.
Importantly, we find that the $^{129}$Xe spins can act as a quantum preamplifier, which converts the external resonant field into an amplified field $|\textbf{B}^n_\text{eff}| \gg |\textbf{B}_\text{ac}|$ without introducing additional noise.

The amplification effect can be quantitatively described by introducing an amplification factor~\cite{jiang2021search,jiang2021floquetspin} 
\begin{equation}
     \label{eff}
    \eta=|\textbf{B}^n_{\textrm{eff}}|/|\textbf{B}_{\textrm{ac}}| = \dfrac{4 \pi}{3} \kappa_0 \gamma_{N} M^{N}_z T_{2}^{N},
\end{equation}
where $\kappa_0 \approx 540$ is the Femi-contact enhancement factor, $\gamma_{N}$ is the gyromagnetic ratio of $^{129}$Xe nucleus;
$M^{N}_z$ is the initial longitudinal magnetization along the polarization axis and
$T_{2}^{N}\approx20\,$s is the transverse relaxation time.
For the $^{129}$Xe-$^{87}$Rb system reported in this work, the amplification factor is measured to be $\eta\approx$ 187.4.
This shows that the spin amplifier can enhance the effect of the exotic field by more than two orders of magnitude, whose magnetic sensitivity reaches $33.9$\,fT/Hz$^{1/2}$ at the resonance frequency, as shown in Fig.\,\ref{figure2}B.
See Methods and Supplementary Section II.C for more details.
In addition, the spin amplifier is only sensitive to a transverse oscillatory field, since fields along the polarization axis cannot induce a measurable transverse magnetization.

Similar resonant techniques based on nuclear magnetic resonance have been recently introduced to search for exotic interactions~\cite{arvanitaki2014resonantly,budker2014proposal,chu2020search}, and most of their experimental demonstrations are still ongoing. 
Similar to the quantum spin amplifier,
these resonant techniques measure exotic fields through measuring the response of nuclear spins with an auxiliary magnetometer.
However, $^{87}$Rb atoms in the spin amplifier act as an embedded magnetometer, enabling $in$-$situ$ measurement and continuous polarization of the $^{129}$Xe atoms. 
A significant advantage offered by $in$-$situ$ measurements is the enhancement of the nuclear resonance signals due to the large Fermi-contact amplification factor.
In addition, since $^{129}$Xe nuclear spins are polarized continuously by spin-exchange collisions with polarized $^{87}$Rb, the spin amplifier can search for exotic fields with a 100\% duty cycle, which is suitable for ultrasensitive continuous wave detection of exotic interactions~\cite{su2021search,wang2022limits}.

In order to resonantly search for the POSS interaction, we modulate the polarization of the spin source, thus modulating the pseudomagnetic field $\textbf{B}_{11}$.
As shown in Extended Data Fig.\,\ref{figure3}A, the exotic signal is modulated by periodically blocking the pump laser beam, with a $50\%$ duty cycle, with an optical chopper. 
Once the chopping frequency matches the resonance frequency (i.e.~$\nu\approx\nu_0\approx10.0\,$Hz), the spin amplifier enables to resonantly search for the exotic field with a sensitivity improvement of about 200 (see Methods for details).

The polarized spin source inevitably generates an usual magnetic field in addition to the exotic field, whose existence poses a challenge for exotic spin-spin interaction searches.
As a result, most experimental studies on spin-spin interactions have focused on meter-scale or kilometer-scale searches~\cite{vasilakis2009limits,hunter2013using}.
In our case, taking advantage of sensitivity anisotropy and small-size magnetic shields, the magnetic field is suppressed by about five orders of magnitude in total, enabling to test the POSS interaction at millimeter-scale distance.
See Methods and Supplementary Section III for details.

\begin{table}[t]
\newcommand{\tabincell}[2]{\begin{tabular}{@{}#1@{}}#2\end{tabular}}
\begin{ruledtabular}
\caption {~~~Summary of calibrated parameters and systematic errors. The corrections to $f_{11}$ for $\lambda =0.1$\,m are listed.} 
\label{table1}
\renewcommand{\arraystretch}{1.2}
\begin{tabular}{l c c}   
Parameter & Value & $\Delta f_{11}\left( \times 10^{-22}\right)$ 
  \\[0.1cm]
\hline
Position of spin source $x$ (mm) & $-1.41\pm0.40$ & $<0.01$  \\[0.15cm]
Position of spin source $y$ (mm) & $50.67\pm0.71$ & $\pm0.07$  \\ [0.15cm]
Position of spin source $z$ (mm) & $3.19\pm0.01$ & $<0.01$  \\ [0.15cm]
Num of polarized Rb & $(2.14\pm0.24)\times 10^{14}$ & ${}^{-0.17}_{+0.20} $ \\[0.15cm]
Phase delay $\phi_a$ (deg) & $13.20\pm0.54$ & ${}^{+0.71}_{-0.45}$\\[0.15cm]
Calib. const. $\alpha$ (V/nT) & $1.99^{<0.01}_{-0.17}$ & ${}^{<0.01}_{+0.19}$\\[0.25cm]
Final $f_{11} ( \times 10^{-22})$ & $2.12$
 & $\pm 5.87 \ (\text{stat}) $\\

$(\lambda =0.1\,\text{m})$ &  & $\pm 0.77 \ (\text{syst})
$ 
\end{tabular} 
\end{ruledtabular}
\end{table}

The experimental search for the POSS interaction consists in calibration experiments and search experiments.
We perform calibration experiments to determine the required parameters, which are summarized in Table\,\ref{table1}.
Calibrated parameters are used to obtain a reference signal of the pseudomagnetic field for subsequent data analysis (see Methods and Supplementary Section IV.A). 
During search experiments, data are collected in records of one hour.
Between every two records, we re-optimized the magnetic sensitivity of the spin amplifier.
The total duration of search data is 24 hours.


\begin{figure}[t]  
	\makeatletter
	\def\@captype{figure}
	\makeatother
	\includegraphics[scale=1.03]{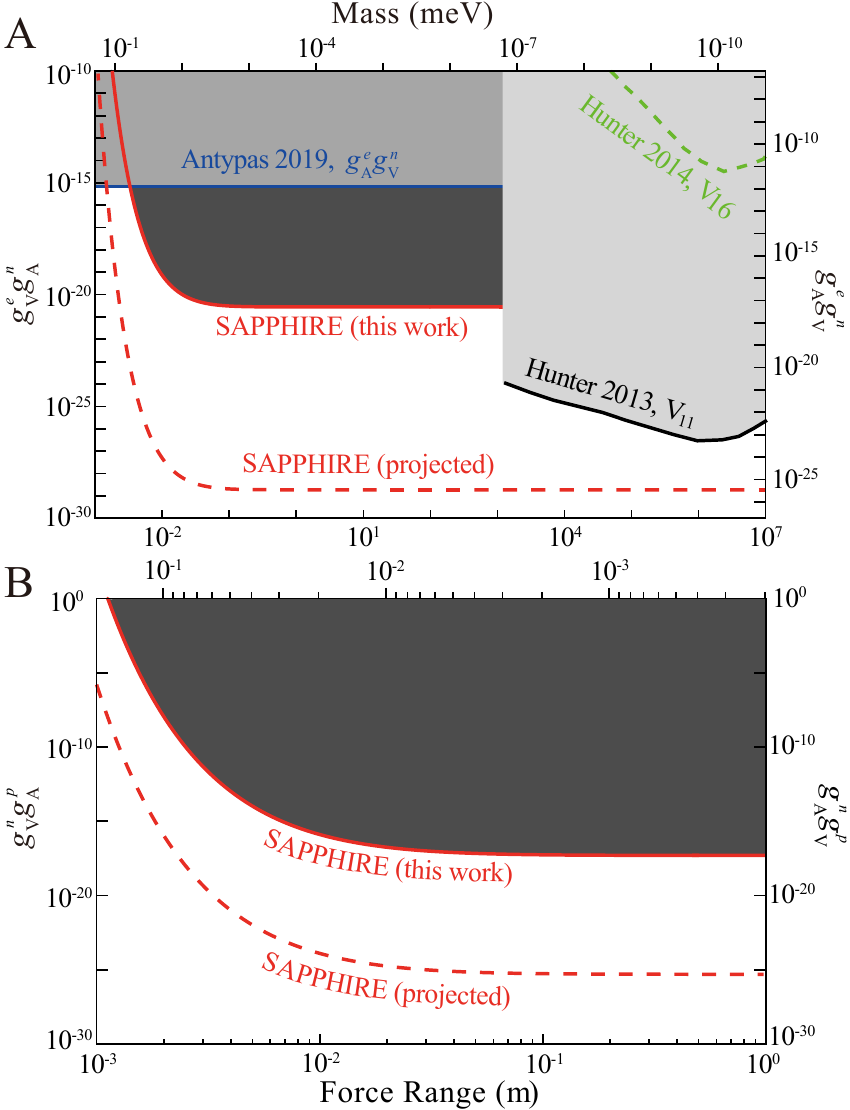}
	\caption{Constraints on POSS coupling strengths.
	The red solid line represents the experimental limits at the 95$\%$ confidence level as a function of the force range $\lambda$ and the boson mass (top axis).
	The red dashed line is projected limits obtained by combining the proposal $^3$He-K spin amplifier with a highly polarized solid-state spin source.
	(A) The vertical axes on the left and right represent the constraints on $g^{e}_Vg^{n}_A$ and $g^{e}_Ag^{n}_V$, respectively. 
	The black solid~\cite{hunter2013using} and green dashed~\cite{hunter2014using} lines indicate the constraints from the measurement of $V_{11}$ and $V_{16}$, respectively, using polarized electrons in the Earth.
	The blue line stands for the constraints on $g^e_Ag^n_V$ from combining results of the Cs and Yb parity-violation experiments~\cite{wood1997measurement,antypas2019isotopic}. 
	(B) The vertical axis on the left and right represent the limits on $g^{n}_Vg^{p}_A$ and $g^{n}_Ag^{p}_V$, respectively. }
	\label{figure5}
\end{figure}

Using the search data, we can constrain the gauge coupling of $Z'$ bosons as mediators of the parity-odd spin-spin interaction~\cite{dobrescu2006spin}, whose existence arises in numerous Standard-Model extensions such as supersymmetric theories and general hidden portal models~\cite{kaneta2017portal}.
The POSS interaction can be induced by exchange of a $Z'$ boson following the effective Lagrangian as depicted by the Feynman diagrams in Fig.\,\ref{figure1}~[\onlinecite{dobrescu2006spin},\onlinecite{fadeev2019revisiting}]
\begin{flalign}
      \mathcal{L}_{Z'}=Z'_{\mu}\sum_{\psi=e,n}\left[\overline{\psi}\gamma^{\mu}(g^\psi_V+g^\psi_A\gamma_5)\psi\right]\, ,
\end{flalign}
where $\psi$ denotes fermion field (electron or neutron), $g^{\psi}_{A} (g^{\psi}_{V})$ stands for the axial (vector) coupling to the fermion $\psi$, $\gamma^\mu$ and $\gamma_5=i\gamma^0\gamma^1\gamma^2\gamma^3$ are Dirac matrices.
The dimensionless coefficient $f_{11}$ related to couplings constants $g^{\psi}_{A,V}$ following the notation of Eq.\,(5.29) in Ref.\,[\onlinecite{dobrescu2006spin}] is given as
\begin{flalign}
     f^{en}_{11}=\frac{1}{2}g^{e}_Vg^{n}_A+\frac{m_e}{2m_n}g^{e}_Ag^{n}_V,
\end{flalign}
where $^{129}$Xe and $^{87}$Rb respectively contribute the neutrons and electrons in our experiment.
Here the contribution of nucleon in $^{87}$Rb is ignored (it is discussed below).

Figure\,\ref{figure5}A shows the experimental limits on the POSS couplings between electrons and neutrons.
For the force range $\lambda>10^3$\,m, the most stringent constraints were set by Hunter et al.~\cite{hunter2013using} as shown with the black line, where the polarized electrons of the Earth were used as the spin source.
However, the work did not constrain the couplings for $\lambda<10^3$\,m,
as their limits are exponentially suppressed at short distances because the relative contribution to polarized electrons from the Earth's outermost crust is only $2\%$~[\onlinecite{hunter2013using}].
For the force range $\lambda<3\times10^{-3}$\,m, the most stringent limits on $g^{e}_Ag^{n}_V$ (blue line) were set from atomic parity-violation tests by detecting the induced electric-dipole transition~\cite{antypas2019isotopic}.
Our searches rely on centimeter-scale detection technique, bridging the gap between atomic-scale and Earth-scale experiments.
In consequence, our result yields the most stringent constraints for $3\times10^{-3}\,\text{m}<\lambda<10^3$\,m (red solid line), improving over previous limits by up to five orders of magnitudes.

Our experiment can also constrain the 
POSS coupling constants between nucleons.
We consider valence protons within the Rb nuclei in the spin source.
The valence proton is nearly fully polarized ($\sim90\%$) due to the high polarization of $^{87}$Rb.
The proton-neutron coupling strength $f_{11}$ can be redefined as
\begin{flalign}
     f^{np}_{11}=\frac{m_e}{2m_p}g^{n}_Ag^{p}_V+\frac{m_e}{2m_n}g^{n}_Vg^{p}_A,
\end{flalign}
where the coupling of $^{87}$Rb electrons is ignored. 
Therefore, our work also provides new experimental limits on $g^{n}_Ag^{p}_V$ (and $g^{n}_Vg^{p}_A$), as shown by the red solid line in Fig.\,\ref{figure5}B.
For the force range below 1\,m, our work sets the best constraints on the POSS coupling and reaches into the hitherto unexplored parameter space to test parity-odd interaction between nucleons.

Astrophysical and cosmological observations can usually be used to constrain the coupling strength between hypothetical bosons and Standard-Model particles.
The coupling constant of spin-0 bosons to electrons and nucleons are tightly constrained by astrophysical limits, such as those from star cooling.
It is the opposite for spin-1 exchanged interactions, where potential loopholes may exist in the relevant theories~\cite{dobrescu2006spin,jain2006evading}.
As a result, our constraints from laboratory direct search for the spin-1 exchanged POSS interaction are the most stringent, where astrophysical limits are relaxed.

There is still room for improvement in the present experiment. We currently modulate spin polarization by chopping circularly polarized pump light; if instead we reverse the direction of the circular polarization, this will yield a factor of two gain in the exotic-field amplitude and, moreover, would allow to better subtract possible systematic effect due to the apparatus imperfections. Such systematics would be further suppressed by adding an additional symmetrically located spin sensor and/or spin source (Fig.\,\ref{figure1}).
Such an arrangement would allow taking full advantage of all possible reversals implied by the form of the POSS interaction, Eq.\,\eqref{v11}.  

Further improvement in experimental sensitivity can be anticipated by replacing the $^{129}$Xe-Rb system in the spin amplifier with a $^3$He-K system~\cite{jiang2021search,su2021search,ji2018new}
and using solid-state spin sources, such as nitrogen-vacancy centers in diamonds~\cite{acosta2009diamonds,schloss2018simultaneous} or pentacene-doped terphenyl crystal~\cite{wu2020invasive,tateishi2014room}.
Combining such $^3$He-K spin amplifier system and solid-state spin sources, the projected constraints are more stringent than our present limits by about eight orders of magnitude (see Methods), as shown by red dashed lines in Fig\,\ref{figure5}.
Over a wide force range, the projected limits would improve on the existing bounds, including the limits from the Earth-geoelectron experiments~\cite{hunter2013using,hunter2014using}.
This makes it possible to discover new physics on $Z'$ bosons with the ultrasensitive SAPPHIRE experiments.

~\

\noindent
\textbf{References}
\bibliographystyle{naturemag}
\bibliography{mainrefs}


~\

\noindent
\textbf{Methods}

\noindent
\textbf{Apparatus.}
Objects with large polarization number density are preferentially chosen as the spin source and placed close enough to the spin sensor for maximizing exotic signals.
As shown in Fig.\,\ref{figure2}A, a 0.58-cm$^3$ cubic glass cell containing polarized $^{87}$Rb vapor with 600\,torr of N$_2$ buffer gas is used as the spin source, and is positioned $50.7$\,mm above the center of the spin-sensor vapor cell.
The spin-source vapor cell is heated to 180$^{\circ}$C to obtain $^{87}$Rb number density ($\approx4.0\times10^{14}$\,cm$^{-3}$).
A laser beam generated with a tapered amplifier system with a power of 0.34\,W is tuned to the D1 transition of $^{87}$Rb and delivered to the spin-source vapor cell along $\hat{z}$ with a single-mode fiber.
The laser beam is circularly polarized and expanded to pump all the $^{87}$Rb in the spin-source cell to an average electron polarization of $>0.9$ and the corresponding number of polarized $^{87}$Rb electron spins is about $(2.1\pm0.2)\times10^{14}$.
Here, the electron polarization along $\hat{z}$ defined as $P_z=2\langle S_z\rangle$ is determined by numerical calculations, where all the required parameters (such as $^{87}$Rb number density and pump laser power etc.) are experimentally calibrated.
The uncertainty of the electron polarization is estimated from the uncertainties of the relevant parameters (see Supplementary Section IV.A for details).
The spin sensor and the spin source are both enclosed within a single-layer magnetic shield (withing the five-layer shield).

In contrast to our recent work~\cite{su2021search} using an unpolarized source for detecting spin-mass velocity-dependent interactions, the polarized spin source in this work inevitably generates a real magnetic field in addition to the exotic field.
The existence of this magnetic fields poses a challenge for spin-spin interaction searches.
Large-size magnetic shields~\cite{vasilakis2009limits} or long-range detection~\cite{hunter2013using} was usually used in the previous search experiments, where the spin source was far from the sensor.
This sets a bottleneck for the sensitivity for short force ranges ($\sim1\,$cm) due to the exponential fall-off of the signal.
As a result, most experimental studies on spin-spin interactions have focused on meter-scale searches, corresponding to mediator masses less than $10^{-4}\,$meV.
In our case, a significant dipole magnetic field of about 1.5\,pT would be produced by the polarized $^{87}$Rb spins acting on the spin sensor if the source and the sensor were not enclosed in additional shielding. 
We use single-layer $\mu$-metal shielding for both the source and the sensor with the total combined shielding factor experimentally determined to be about $10^4$.
In addition, we arrange the spin source directly above the spin sensor (Fig.\,\ref{figure2}A), where the dipole magnetic field is oriented along the insensitive axis of the spin amplifier. 
The component of the magnetic field experienced by the spin sensor along the sensitive directions is estimated to be about $10$\,aT (see Supplementary Section III), which can be negligible with respect to the 0.1\,fT magnetic field detectable by the spin amplifier for twenty-four hours of operation. 
Taking advantage of sensitivity anisotropy and small-size magnetic shields, the magnetic field is suppressed by about five orders of magnitude in total, enabling to test the POSS interaction at short distance, down to about the mm range. 

~\

\noindent
\textbf{Amplification factor and magnetic-field sensitivity.}
We calibrate the amplification factor and magnetic sensitivity of the quantum spin amplifier which is important for determining the sensitivity to the POSS interaction.
By comparing the response to external resonant and off-resonant fields, the amplification factor is measured to be $\eta\approx$ 187.4.
This shows that the spin amplifier can enhance the effect of the exotic field by more than two orders of magnitude. 
By scanning the frequency of the applied oscillatory field, we obtain the frequency dependence of the sensitivity to magnetic fields along $\hat{x}$ and $\hat{z}$, respectively depicted in Fig.\,\ref{figure2}B.
The $^{129}$Xe Lamor frequency is tuned to $\nu_0\approx10.0$\,Hz by setting the constant magnetic field $B_0\hat{z}$ at 847\,nT to avoid low-frequency noise. 
The optimal magnetic sensitivity is $33.9$\,fT/Hz$^{1/2}$ at the resonance frequency, with the oscillatory field applied along $\hat{x}$, while the off-resonance sensitivity is only $6.4$\,pT/Hz$^{1/2}$.
The spin amplifier is only sensitive to a transverse oscillatory field, since fields along the polarization axis cannot induce a measurable transverse magnetization.
As a result, the residual sensitivity to $\hat{z}$-fields (owing to the Rb magnetometer) of ($\approx257.5\,\text{pT/Hz}^{1/2}$) is about two orders of magnitude worse than that for $\hat{x}$ fields.
Taking advantage of the sensitivity anisotropy of the spin amplifier, it is possible to distinguish spurious magnetic fields from the pseudomagnetic field.

~\

\noindent
\textbf{Modulation technique and resonant search.}
In order to resonantly search for the exotic POSS interaction, we modulate the polarization of the spin source, thus modulating the pseudomagnetic field $\textbf{B}_{11}$.
As shown in Extended Data Fig.\,\ref{figure3}A, the exotic signal is modulated by periodically blocking the pump laser beam, with a $50\%$ duty cycle, with an optical chopper. 
The modulated pseudomagnetic field as a function of time can be represented as
\begin{flalign}
     \!\textbf{B}_{11}(t)&\!=\!\textbf{B}_{11}^0\cdot\left\{\frac{1}{2}+ \frac{\mathcal{S}\left[\text{sin}(2\pi\nu t+\phi)\right]}{2}\right\} ,
     \nonumber
     \\
     &=\sum \limits_{N=\text{odd}} \textbf{B}^{(N)}_{11} \frac{\sin(2 \pi N \nu t+\phi_N)}{2}+C,
\label{B11_3}
\end{flalign}
where $\textbf{B}_{11}^0$ is the pseudomagnetic field produced by the spin source under pumping; $\mathcal{S}$ stands for the signum function; $\phi$ is the initial phase of the exotic signal; $|\textbf{B}^{(N)}_{11}|=|\textbf{B}_{11}^0|\cdot4/(\pi N)$ is the peak-to-peak amplitude of the $N$th harmonic, $\phi_N$ is the corresponding phase and $C=\textbf{B}_{11}^0/2$ is the DC component of the signal.
Here, we ignore the rising edge and falling edge of the exotic square-wave signals because the transient of the source polarization which is faster than the relaxation time of electrons ($\sim$1\,ms) is much shorter than the pumping period ($\sim$100\,ms).
Once the optical chopping frequency matches the resonance frequency (i.e.~$\nu\approx\nu_0\approx10.0\,$Hz), the spin amplifier enables to resonantly search for the exotic field with a sensitivity improvement of about 200.

We experimentally test the sensitivity of the resonant measurement by applying a  square-wave (real) field to the spin amplifier to simulate the modulated exotic field with an amplitude of $|\textbf{B}_{11}^0|$, whose Fourier-transform spectrum contains only odd harmonics (Extended Data Fig.\,\ref{figure3}B).
As the primary component of simulated exotic fields, the first harmonic with peak-to-peak amplitude of $|\textbf{B}^{(1)}_{11}|$ is about 1.27 times larger than the signal $\textbf{B}_{11}^0$. 
The resonance frequency of spin amplifier is tuned to match the frequency of the modulated field.
Only the first harmonic of the field is amplified by the spin sensor and the effects of other harmonics are negligible (Extended Data Fig.\,\ref{figure3}C).

~\

\noindent
\textbf{Data analysis.}
In data analysis, the exotic reference signal of pseudomagnetic fields must be obtained first in order to extract weak signals from noisy environments~\cite{ji2018new,su2021search,wang2022limits}.
Because only the first harmonic is amplified,
we obtain the reference signal as $|\textbf{B}^{(1)}_{11}|/2\cdot\sin(2 \pi \nu t+\phi_1-\phi_a)$, where the frequency is equal to the modulation frequency $\nu=10.0$\,Hz and the phase is obtained as $\phi_1-\phi_a$ due to the existence of phase delay.
The amplitude of reference signal $|\textbf{B}^{(1)}_{11}|/2$ can be determined by the input exotic field $\textbf{B}_{11}^{0}$ [see Eq.\,(\ref{v11})]
\begin{flalign}
     \!\textbf{B}_{11}^{0}(f_{11})&\!=\!\frac{-f_{11}}{4\pi m_{e} \mu_{\textrm{Xe}}}\! \int_{V}\!\rho(\bm{r})(\!\hat{\sigma}_e\!\times\!\hat{r})\left(\!\frac{1}{\lambda r}\!+\!\frac{1}{r^{2}}\right)\!e^{-r/\lambda}
      d\bm{r},
\end{flalign}
where $V$ represents the integration space of the internal volume for the spin-source cell and $\rho(\bm{r})$ is the number density of the polarized electron spins.
Here the volume shape and polarization gradient of the spin source should be considered (see Supplementary Section IV.B for details).

To obtain the constraints on coupling strength $f_{11}$, a ``lock-in'' analysis method~\cite{su2021search,wang2022limits,ji2018new} is introduced to extract exotic signals.
This method derives the wanted reference signal with a certain frequency and phase from the experimental signal $S(t)$ and determines its amplitude. 
In order to obtain statistical errors of the coupling strength within each one-hour data set, the experimental signal $S(t)$ is separated into segments $S(t)_i$ with a single period $T=0.1$\,s.
The coupling strength $f^{i}_{11}$ for every period is calculated as 
\begin{equation}
    f^{i}_{11}= \frac{\pi}{2\alpha |\textbf{B}^0_{11}(1)|} \frac{\int^{T}_0 \sin{(2 \pi \nu t+\phi_1-\phi_a)} S(t)_i\,d t}{\int^{T}_0 \sin^2{(2 \pi \nu t+\phi_1-\phi_a)}\,d t}, 
\end{equation}
where $\alpha$ is the calibration constant which converts the output signal (V) of the spin amplifier to the exotic field (nT) and $\textbf{B}^0_{11}(1)$ is the input pseudomagnetic field corresponding to $f_{11}=1$. 
The experimentally measured coupling strengths $f^{i}_{11}$ are fitted with a Gaussian distribution to determine the mean value and statistical error for each one-hour data set, as shown in the inset of Extended Data Fig.\,\ref{figure4}.  
For the force range $\lambda=0.1$\,m, the coupling strength for total 24\,h is obtained as $f_{11}\approx(2.1\pm5.9_{\text{stat}})\times10^{-22}$ with a weighted reduced $\chi^2\approx1.34$.
The corresponding systematic errors are summarized in the third column of Table\,1, in which the major systematic error comes from the phase delay of the spin amplifier.
By combining all the systematic errors in quadrature, which are assumed to be independent of each other, the overall systematic uncertainty is derived.
Accordingly, for the force range $\lambda=0.1$\,m, we quote the final coupling strength as $f_{11}\approx(2.1\pm5.9_{\text{stat}}\pm0.8_{\text{syst}})\times10^{-22}$ and the constraint can be determined as $1.5\times10^{-21}$ at the $95\%$ confidence (see Supplementary Section V).
By varying the force range $\lambda$ and repeating the above process, constraints for the entire explored force range were obtained.

~\

\noindent
\textbf{Projected improvement in SAPPHIRE.}
Further improvement in experimental sensitivity to the parity-odd spin-spin interactions can be anticipated by replacing the Xe-Rb system in the spin amplifier with a $^3$He-K system~\cite{jiang2021search,su2021search,ji2018new}.
As seen in Eq.\,(\ref{eff}), to realize larger amplification, a prerequisite is to achieve long transverse relaxation time and high polarized spin density of noble gas.
Specifically, the predicted amplification factor of a $^{3}$He-K system can be better than the current one by at least two orders of magnitude~\cite{vasilakis2009limits}. 
In addition to the amplification factor, K magnetometers have higher magnetic sensitivity (a few fT/Hz$^{1/2}$) than Rb magnetometers~\cite{vasilakis2009limits}.
Overall, the projected magnetic sensitivity of $^{3}$He-K spin amplifier can be improved by four orders of magnitude over the current spin amplifier.

The number density of polarized spins in the source can be greatly increased by using solid-state spin sources, such as nitrogen-vacancy centers in diamonds~\cite{acosta2009diamonds,schloss2018simultaneous} and pentacene-doped terphenyl crystal~\cite{wu2020invasive,tateishi2014room}.
Near-unity spin polarization can be realized with optical pumping.
For example, a crystal of $p$-terphenyl doped with $0.1\,\text{mol}\%$ fully deuterated pentacene can be a good candidate for high-polarization spin sources, with polarized-electron density reaching  $\sim1.5\times10^{18}$\,$\text{cm}^{-3}$ under optical pumping~\cite{wu2020invasive}.
The high nucleon polarization of synthesized deuterated $p$-terphenyl-2$'$,3$'$,5$'$,6$'$-$d_4$ ($\sim34\%$) can be created by dynamic nuclear polarization (DNP) in 0.4\,T at room temperature, which transfers the polarization from the electrons of deuterated pentacene $0.05\,\text{mol}\%$ to the protons of terphenyl, with the total number of polarized protons achieved exceeding $3\times10^{18}$\,[\onlinecite{tateishi2014room}].
Overall, using optically pumped pentacene-doped terphenyl crystals, one could increase the strength of exotic field by about four orders of magnitude.
Combining the enhanced spin amplifier based on $^3$He-K system discussed above and highly polarized solid-state spin sources, the projected constraints are more stringent than our present limits by about eight orders of magnitude (red ashed lines in Fig\,\ref{figure5}).

\setcounter{footnote}{50} 

~\

\noindent
\textbf{Data availability}.

\noindent
Source data are provided with this paper. All other data that support the plots in this paper and other findings of this study are available from the corresponding author upon reasonable request.

~\

\noindent
\textbf{Code availability}.

\noindent
The code that supports the plots in this paper is available from the corresponding author upon reasonable request.

~\

\noindent
\textbf{Acknowledgements}.

\noindent
We thank Hao Wu and Dong Sheng for valuable discussions.
This work was supported by National Key Research and Development Program of China (grant no. 2018YFA0306600), National Natural Science Foundation of China (grants nos. 11661161018, 11927811, 12004371), Anhui Initiative in Quantum Information Technologies (grant no. AHY050000),
USTC Research Funds of the Double First-Class Initiative (grant no. YD3540002002) and the Deutsche Forschungsgemeinschaft (DFG) - Project ID 423116110. This work was also supported by the Cluster of Excellence ``Precision Physics, Fundamental Interactions, and Structure of Matter'' (PRISMA+ EXC 2118/1) funded by the German Research Foundation (DFG) within the German Excellence Strategy (Project ID 39083149).

~\

\noindent
\textbf{Author contributions}.

\noindent
Y.H.W. and Y.H. designed experimental protocols, performed experiments, analyzed the data and wrote the manuscript.
C.G., X.K., and Y.S.Q. performed experiments and edited the manuscript.
H.W.S., W.J., and D.D.H. analyzed the data and edited the manuscript.
M.J., X.H.P., and D.B. proposed the experimental concept, designed experimental protocols and proofread and edited the manuscript.
All authors contributed with discussions and checking the manuscript.

~\

\noindent
\textbf{Competing interests}.

\noindent
The authors declare no competing interests.

\newpage

\begin{figure*}[ht]  
	\makeatletter
	\def\@captype{figure}
	\makeatother
	\setcounter{figure}{0} 
	\includegraphics[scale=1.05]{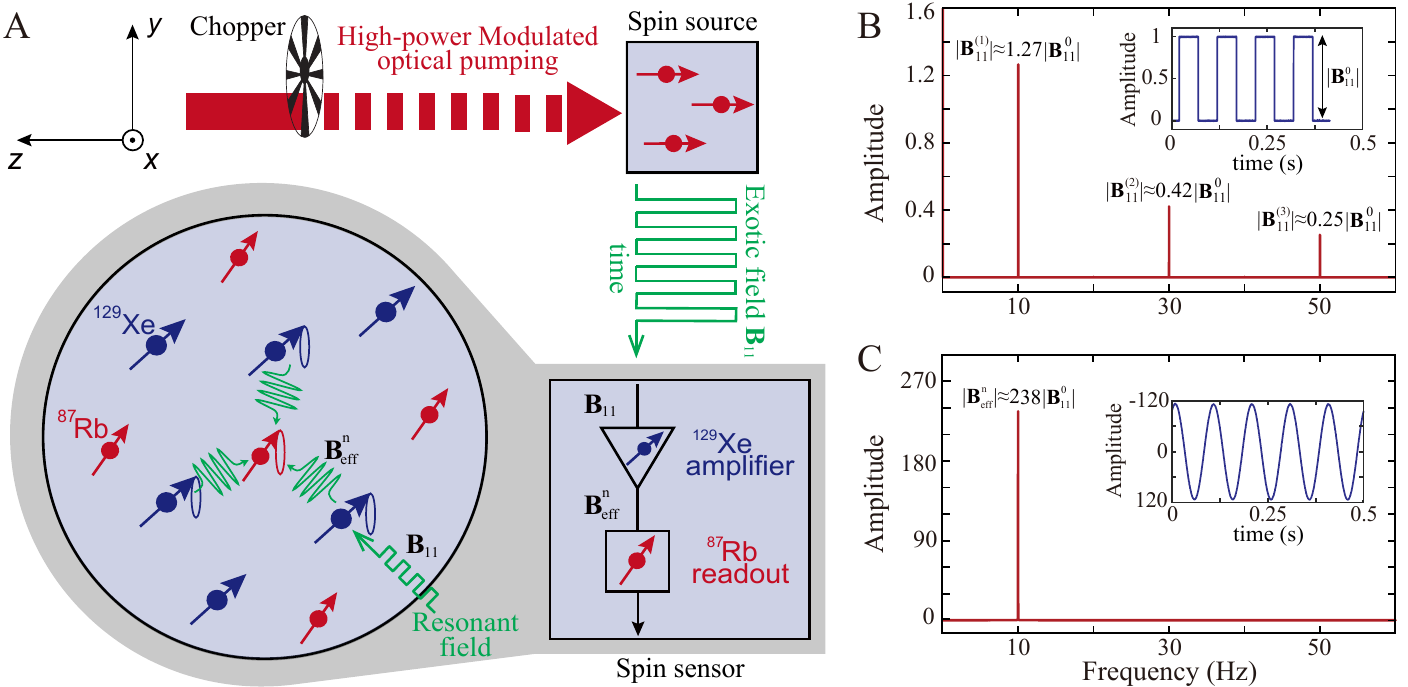}
	\renewcommand{\figurename}{Extended Data Fig}
	\caption{Demonstration of resonance-sensitive search for exotic signal based on spin amplifier.
	(A) Under a high-power ($>$0.34\,W) modulated optical pumping along $\hat{z}$, the spin source generates a $\hat{x}$-directed square-wave pseudomagnetic field on the spin senor with 50$\%$ duty cycle at a frequency of $\nu\approx10.0$\,Hz.
	The exotic signal $\textbf{B}_{11}$ is enhanced by the $^{129}$Xe atoms, yielding an effective field $\textbf{B}^n_{\text{eff}}$ read out by $^{87}$Rb atoms.
	(B) Ratios of the peak-to-peak amplitude of square-wave's Nth harmonic relative to the amplitude of original signal $|\textbf{B}^0_{11}|$.
	Right top inset: Experimental measured time-domain square-wave signal of simulated exotic field $\textbf{B}_{11}$. 
	(C) Experimental response in frequency domain of the spin amplifier to the square-wave magnetic field.
	Only the first harmonic of the input exotic signal $\textbf{B}^{(1)}_{11}$ can be amplified and detected.
	}
	\label{figure3}
\end{figure*}

\begin{figure*}[t]  
	\makeatletter
	\def\@captype{figure}
	\makeatother
	\includegraphics[scale=1.03]{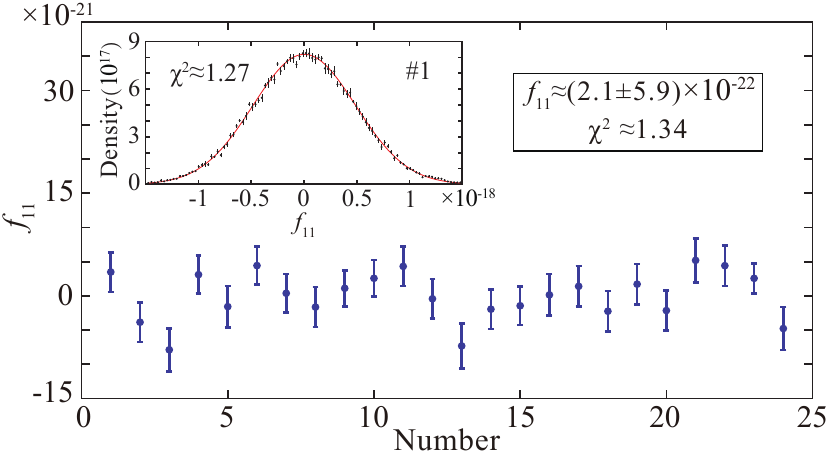}
	\renewcommand{\figurename}{Extended Data Fig}
	\caption{Experimental coupling strength $f_{11}$ for the force range $\lambda=0.1\,$m. Each point and error bar stands for mean value and statistical error for one hour respectively.
	Left top inset: Graph of values for the first-hour record ($\#$1) follows a Gaussian distribution.
	}
	\label{figure4}
\end{figure*}


\end{document}